\documentclass[12pt,preprint]{aastex}

\newcommand{\etal}{{et\thinspace al.} }
\newcommand{\Lya}{{$Ly\alpha$} }

\begin{document}

\title{Searching for $z\simeq 6$ Objects with the $HST$ Advanced Camera for
Surveys: Preliminary Analysis of a Deep Parallel Field}
\author{Haojing Yan, Rogier A. Windhorst and Seth H. Cohen}
\affil{Department of Physics and Astronomy, Arizona State University, Tempe, AZ 85287-1504}
\email{Haojing.Yan@asu.edu, Rogier.Windhorst@asu.edu, Seth.Cohen@asu.edu}

\begin{abstract}

Recent results suggest that $z\simeq 6$ marks the $end$ of the reionization 
era. A large sample of objects at $z\simeq 6$, therefore, will be of enormous 
importance, as it will enable us to observationally determine the exact epoch
of the reionization and the sources that are responsible for it. With the HST 
{\it Advanced Camera for Surveys} (ACS) coming on line, we now have an unique
opportunity to discover a significant number of objects at $z\simeq 6$. The 
pure parallel mode implemented for the {\it Wide Field Camera} (WFC)
has greatly enhanced this ability. We present our preliminary analysis of a 
deep ACS/WFC parallel field at $|b|=74.4^o$. We find 30 plausible $z\simeq 6$
candidates, all of which have $S/N>7$ in the F850LP-band. The major source of
contamination could be faint cool Galactic dwarfs, and we estimated that they
would contribute at most 4 objects to our candidate list. We derived the 
cumulative number density of galaxies at $6.0\leq z\leq 6.5$ as 2.3 
arcmin$^{-2}$ to a limit of 28.0 mag in the F850LP-band, which is slightly
higher than our prediction. If this is not due to an underestimated
contamination rate, it could possibly imply that the faint-end slope of the
$z\simeq 6$ luminosity function is steeper than $\alpha=-1.6$. At the very 
least, our result suggests that galaxies with $L<L^*$ do exist in significant
number at $z\simeq 6$, and that they could be the major sources that
contributed the reionizing photons.

\end{abstract}


\section{Introduction}

  In the last couple of years, great progress has been made in answering the
grand cosmological question of when reionization of the universe began.
It is now generally believed that we have seen {\it the end} of the 
reionization at around $z=6$. Fan \etal (2002) provided a detailed analysis of
the spectra of the SDSS $z\sim 6$ quasars, and argued that the epoch of 
reionization could not be at a redshift much higher than 6. This is also 
consistent with the most recent numerical simulations of cosmological
reionization (e.g., Cen \& McDonald 2002), where it is suggested that the IGM
is likely neutral at $z > 6.5$. On the other hand, Hu \etal (2002) 
reported their discovery of a $z=6.56$ \Lya emitter and therefore suggested 
that the reionization lies beyond $z=6.6$. Nevertheless, the detectability of
\Lya emission line $prior$ to the reionization was investigated by Haiman
(2002), who pointed out that such a line, if broad enough, could still be
observable, and thus the \Lya emitter of Hu \etal (2002) might actually lie
$beyond$ the epoch of the reionization. Very recently, Cen (2002) proposed
a scenario that the Universe was actually reionized twice, which could also
explain the seemingly-paradoxical detection of this $z=6.56$ \Lya emitter.

  The kinds of sources responsible for the reionization is one other question
that remains open. While the strong UV spectrum of a quasar is capable of
ionizing surrounding IGM out to Mpc scales, the paucity of quasars makes them 
unlikely the major contributors to the reionizing background. Based on the 
stringent constraint they obtained for the bright-end slope of quasar 
luminosity function at $z\sim 6$, Fan \etal (2002) pointed out that luminous 
quasars could not provide sufficient photons to keep the universe ionized at
this redshift. However, there is still room for lower luminosity AGNs 
($M_B > -23$ mag) being important reionizing sources. Alternatively, the
reionizing photons could also come from star-forming galaxies, provided that a
sufficient amount of Lyman continuum emission could escape from the galaxies.
While the escape fraction ($f_{esc}$) of such emission is found to be extremely
small in the local universe (Leitherer \etal 1995; Hurwitz \etal 1997), Steidel
\etal (2001) derived $f_{esc}\simeq$ 65\% from the composite spectrum of a
sample of 29 $z\simeq 3.4$ Lyman-break galaxies. Although such a high $f_{esc}$
value is still under debate (see Giallongo \etal 2002), star-forming galaxies cannot
be ruled out as important contributors to the reionizing background. An extreme
scenario has been proposed by Ricotti (2002), who suggested that
globular clusters, which can be formed outside of galaxy disks or bulges and
thus have $f_{esc}$ close to unity, were responsible for the reionization.

   Obviously, we will have to first acquire a large sample of $z\simeq 6$ 
objects before we can settle all these issues. 
The {\it Wide Field Camera}
(WFC) of the {\it Advanced Camera for Surveys} (ACS) on-board the HST provides
an unique opportunity to discover a large number of $z\simeq 6$ objects and 
probe the faint end of the luminosity function (LF). To maximize its scientific
return, a default pure parallel observation mode has been implemented for the
WFC (Sparks \etal 2001). This mode utilizes four filters, namely 
F775W (SDSS-$i$), F850LP (SDSS-$z$), F475W (SDSS-$g$), and F625W (SDSS-$r$),
in order of preference. Whenever there is at least
one orbit of time available for parallel observation, this mode always takes 
images in the F775W and F850LP bands. Given the high throughput and the wide 
field-of-view (FOV) of the WFC, such a strategy makes its parallel data 
extremely useful for selecting objects at $z\simeq 6$ by using the 
{\it drop-out} technique.

    In this letter, we present our preliminary analysis of a field that is the
deepest ACS/WFC parallel observation to date. We describe the data reduction in
\S 2, followed by the $z\simeq 6$ candidate selection in \S 3. The comparison
between the result of this study and the prediction of Yan \etal (2002) is
made in \S 4. We conclude with a brief summary in \S 5. Throughout the paper,
we use the AB magnitude system, and adopt a cosmological model with 
$(\Omega_M,\Omega_\Lambda, H_0)=(0.3,0.7,65)$.

\section{Data and Photometry} 

   The center of this field is $RA=12^h43^m32^s$, $Dec=11^o40'32''$ (J2000),
and it is at a galactic latitude of $|b|=74.4^o$.
The primary observation of the visits is the WFPC2 imaging of NGC4647,
which has a size of 2.9$^{'}\times 2.3^{'}$ and the V-band total magnitude of
11.3 mag. As the border of the WFC is about $5^{'}$ away from that of the WFPC2,
the parallel field is not contaminated by the light from either this galaxy or
its close neighbor, M60. The observations spanned from April 28 to June 19, 2002.
Only the F775W and F850LP filters were used for the parallel imaging. In total,
15 images were taken in the F850LP-band and 27 images were taken in the
F775W-band. The total exposure time in these two bands is 2.65 hours and 4.28
hours, respectively.

   After combining the individual images into final stacks, we used 
$SExtractor$ of Bertin \& Arnouts (1996) to perform matched-aperture
photometry by invoking its double-input mode. The F850LP stack was used for 
extracting sources and defining apertures, and the magnitude of each detected
source was measured on the F850LP stack and the F775W stack independently,
but with a same aperture. For source detection, we used a $5\times 5$ 
Gaussian smoothing kernel with the FWHM of 2.0 pixels, which is approximately
the same as the FWHM of a point source PSF on both stacks. The detection
threshold was set to 1.8 $\sigma$, and at least 4 connected pixels above this
threshold were required
for a source to be included. We used total magnitude (corresponding to the
{\it mag-auto} option in SExtractor) for the photometry.
The two catalogs were then merged into a master catalog, which we shall
refer to as the ``matched catalog".  
We adopted the zeropoints
used by the Great Observatories Origins Deep Survey (Dickinson \& Giavalisco
2002) team for their HST Treasury program utilizing the same instrument, 
where they have derived the zeropoints as one electron corresponding to 
$m_{775W}=25.656$ mag and $m_{850LP}=24.916$ mag.

   We assessed the F850LP-band survey limit as the following.
The representative error ($\Delta m$) reported by $SExtractor$
was used to calculate the S/N of each extracted source, using the simple 
relation of $\Delta m = 1.0857/(S/N)$. Only the sources with $S/N \geq 5$ were
included in the assessment. From the source count histogram, we estimated that
the survey was 100\% complete to $m_{850LP}\leq 28.0$ mag. To estimate the faintest
level that the F775W-band achieved, we ran $SExtractor$ independently on the
F775W stack with the detection threshold lowered to 1\,$\sigma$. We counted
the number of detected objects regardless of their $S/N$, and found that the
number count histogram reaches its peak value at 30.0 mag. Thus, for an object
that is not detected in the F775W-band, it must be fainter than 30 mag in
this band.  We will use this result in \S 3.2 below.

\section{Selecting $z\geq 6$ Objects}

   We selected $z\geq 6$ objects via the {\it drop-out} technique (e.g.,
Steidel \etal 1995) that identifies the Lyman-break signature their spectral
energy distributions (SED's). At $z\geq 4$, this signature,
which is mainly due to the cosmic intervening H I absorption (Madau 1995),
occurs at rest-frame wavelength of 1216\,{\AA}. At $z\simeq 6$, this 
signature moves out of the F775W-band and into the F850LP-band.
As the area enclosed by the F850LP-band system response curve drops to half of
the total value at around 9200\,{\AA}, this passband losses its efficiency at
this wavelength. Therefore, the combination of these two filters is effective
in identifying the Lyman-break in the redshift range of $6.0\leq z \leq 6.5$.

\subsection{Drop-out Selection}

   We define an object as a ``F775W drop-out" if it is significantly detected
in the F850LP-band but is not visible in the F775W-band. To be specific, such
a source should be flagged as not detected in the F775W-band in the matched
catalog, and it should have reported photometric error smaller than 0.15 mag,
or equivalently, have S/N larger than 7.2, in the F850LP-band. The later
constraint was rather conservative, and it was applied because at this stage we
were more concerned with the reliability of the selection than its
completeness at the low brightness level. At this S/N level, the survey in the
F850LP-band is 100\% complete to 27.5 mag.

   Such criteria resulted in a candidate list of 114 objects. All these sources
were then visually examined in both bands to make sure that 1) it was a real
detection in the F850LP-band and 2) it was not seen in the F775W-band. This 
refining procedure rejected 84 objects from the list and only 30 plausible
candidates remained. We found that there were a variety of reasons that gave
rise to false candidates. Besides those causes commonly seen in CCD
imaging (e.g., background anomaly close to the field edges due to dithering),
one other important cause is the correlated background noise that is due to the
geometric correction, which accounts for about 40\% of the 84 false
candidates. 

   While it maps the off-axis ACS images back to the proper geometry, the
geometric correction makes the mapped pixels non-independent. As a result, the
background noise in the mapped pixels is spatially correlated, and such a
correlation shows up as weak, web-like structures in the background
(visible in Fig. 1). The effect of these
structures is two-fold.  When the object search is pushed to very faint
threshold in the F850LP-band, some of these structures could be picked up as
local maximum and result in false detections in this band. On the other hand,
if a faint but real object is too close to such structures in the F775W-band,
it could be missed by the source detection algorithm. The source detection
parameters that we used to generate the matched catalog were optimized to make
the detection as complete as possible while keeping the number of false objects
manageable. The visual examination serves as a second safeguard procedure to
keep the candidate list clean.

   To independently check the reliability of these 30 candidates, we performed
a simulation to test whether these objects could be recovered if they were at
different places on the F850LP stack. For a given candidate, a $13\times 13$
pixel image stamp centered on it was copied from the F850LP stack. After
subtracting a constant sky background, the stamp was put at 1,000 random 
positions on the F850LP stack, and thus generated 1,000 artificial objects that
have the same photometric properties of their prototype. $SExtractor$ was then
run with the same parameter setting and the number of recovered artificial 
objects was counted. The simulation was done for each of the 30 candidates,
and we found that the median recovering rate was 88\%. The recovering rate
essentially remains at this level for the objects that are brighter than 28.0
mag, and drops slightly to 79\% beyond this limit. 
Thus our candidates are deemed reliable.

   Fig. 1 shows the F775W and the F850LP images of six candidates that are
randomly chosen from the 30 sources. The images have been smoothed by a 
$5\times 5$ boxcar to enhance the detected sources. The F850LP-band magnitude
of the 30 sources ranges from 26.8 to 28.3 mag, and only three of these sources
are fainter than 28.0 mag. The median magnitude of these candidates is 27.4 mag.

\subsection{Possible Sources of Contamination}

   We visually examined the locations of these 30 candidates on each of the 15
individual F850LP-band images, and found no transient event on these locations.
Therefore, any possible source of contamination must be of non-transient nature.
There are three types of sources whose SED could mimic the Lyman-break
signature at $z\simeq 6$, namely, strong emission-line galaxies at low
redshift, elliptical galaxies, and cool dwarfs in our Galaxy. Given the sharp
cut-off of the F775W-band in the red and our stringent selection criteria, we
argue that none of these sources is likely to seriously contaminate our
candidates: 

   1. {\it Low-z emission-line galaxies} --- Such galaxies should have very
strong emission lines in the F850LP-band to be included by our selection
criteria. The observed equivalent width $W_{obs}$ of such an emission line can
be estimated by using $W_{obs}\simeq (\widetilde{F_{\nu}}/f_{\nu c})D$, where 
$D$ is the width of the
F850LP passband, $\widetilde{F_{\nu}}$ is the average flux measured in the 
F850LP-band, and $f_{\nu c}$ is the continuum flux estimated from the 
F775W-band. Since $D$ is about 1390\,{\AA} and a conservative lower limit of
$\widetilde{F_{\nu}}/f_{\nu c}$ is 4, we get $W_{obs} > 5000$\,{\AA}. 
Considering that the system response of the F850LP-band drops to half of its
peak value at around 8300\,{\AA} and 9700\,{\AA} in the blue and the red, 
respectively, the emission lines that are possible to produce the detected flux
are $H{\alpha}$(6563\,\AA) at $0.26\leq z\leq 0.48$, $[O III]$(5007\,\AA) at 
$0.66\leq z\leq 0.94$ and $[O II]$(3727\,\AA) at $1.23\leq z\leq 1.60$. 
Therefore, the smallest possible rest-frame equivalent width is still 
enormously high, $W=W_{obs}/(1+1.60)>2200$\,{\AA}. Such emission-line galaxies
should be very rare, if there is reason to believe that they exist at all.
Hence, it is very unlikely that emission-line galaxies would contaminate our
candidates.

   2. {\it Elliptical galaxies} --- Although the 4000\,{\AA} break in the
SED of an elliptical galaxy at $1.0\leq z\leq 1.5$ occurs in between the F775W
and the F850LP bands, this break is not as steep as the Lyman-break at
$z\simeq 6$. The sharp cut-off of the F775W system response in the red further
helps to discriminate between these two features. Convolving a typical SED of
E/S0 galaxies (e.g., Coleman \etal 1984) at $1.0\leq z\leq 1.5$ with the WFC 
system responses gives $m_{775W}-m_{850LP} \simeq 1.0$ mag, while the same
synthesis for a model $z\simeq 6$ galaxy gives
$m_{775W}-m_{850LP}\geq 2.0$ mag. Given that any object with $m_{775W}<30.0$
mag should be detected in the F775W-band (see \S 2), all of our candidates
have $m_{775W}-m_{850LP}>1.5$ mag, among which 27 objects have
$m_{775W}-m_{850LP}>2.0$ mag. Therefore, we argue that the possible
contamination rate due to elliptical galaxies is three objects at most.

   3. {\it Cool dwarfs} --- L and T type dwarfs are known to create large color
discrepancy in the passbands similar to those used in this study, and it is
very difficult to distinguish them without IR photometry (e.g. Fan \etal 2000).
Unfortunately, we do not have any knowledge about their spatial distribution,
and we have to heavily rely on very rough assumptions to estimate their
contamination to our sample. Given the faintness of our candidates, it is
implausible for luminous dwarfs to enter our selection, as that would put their
distances far beyond the thickness of the Galaxy. The least luminous L dwarfs 
currently detected have absolute magnitudes at $M_I\sim 20$ mag level
(e.g. Dobbie \etal 2002). For such objects to remain undetected in the 
F775W-band, they would have to be at a distance of 0.6 $kpc$ or further.
Assuming an outer bound of 1.5 $kpc$ from the Sun, the volume that our 
field-of-view samples is approximately 880 $pc^3$. If we assume an IMF of the
form $dN/dM \propto m^{-\alpha}$ with $\alpha \simeq 1$ and use the model
of Haywood \& Jordi (2002), the local number density of dwarfs is approximately
of $0.04\,pc^{-3}$. Using a scale height of 0.3 $kpc$ for the Galaxy (e.g.
Liu \etal 2002), the
number density extrapolated to a distance of 0.6 $kpc$ and beyond will drop by
at least a factor of ten. Thus we estimate that the contamination due to cool
dwarfs is about 3--4 objects at most.

\section{Comparison to Prediction}

   Among the 30 plausible candidates, 27 have $m_{850LP}<28.0$ mag. The 
remaining three objects fall in the 28.0--28.5 mag bin where the survey begins
to be severely affected by incompleteness, and, as discussed above, they are
also the only three objects that could possibly be elliptical galaxies
(see \S 3.2). These three objects are not included in the discussion below.

   Taking into account that we discarded the field edges and the
gap between the two CCD chips, the effective survey area is approximately 10
arcmin$^2$. If we assume that cool Galactic dwarfs contribute
4 contaminators (\S 3.2), the number density at
$6.0\leq z\leq 6.5$ is 2.3 arcmin$^{-2}$. This result indicates that $L<L^{*}$
galaxies do exist in large number at $z\simeq 6$ as expected.
We have predicted the number density of galaxies at $5.5\leq z\leq 6.5$
in Yan \etal (2002) by extrapolating the LF measured at $z\simeq 3$ to 
$z\simeq 6$ and using the observational limits in the
HDF-N as the normalization. The high-normalization case of our prediction
gives a cumulative number
density of 3.67 arcmin$^{-2}$ to 28.0 mag. The co-moving volume at 
$6.0\leq z\leq 6.5$ is about 48.7\% of that at $5.5\leq z\leq 6.5$ in our 
adopted model cosmology. Therefore, the number density that Yan \etal (2002)
would predict for the $6.0\leq z\leq 6.5$ range is 1.8 arcmin$^{-2}$.
Fig. 2 shows the cumulative number density prediction adapted from Yan \etal
(2002), with the result from this study added. Note that
the observed number density inferred from our candidates is slightly higher
than the predicted value. While the survey in the F850LP-band is 
complete to 28.0 mag in the matched catalog, the candidate selection is done at
a much higher brightness level ($S/N\geq 7.2$) and is complete to 27.5 mag. 
That means the true number density at 28.0 mag could be even larger.

   This discrepancy could be due to the possibility that we
underestimated the contamination rate of cool dwarfs.
However, the higher number density could also be real. If this is the case,
it implies that the actual LF is slightly different from that in Yan \etal
(2002). Indeed, if we increase the normalization in Yan \etal (2002) by 50\%,
this higher value can be explained without violating any other existing 
constraints to the LF. Or alternatively, if we change the assumed faint-end
slope from $\alpha=-1.6$ to $-2$, this higher value can also be explained.
The later possibility is of particular interest, as a steeper faint-end slope
in the LF is what expected from the reionizing photon budget argument if
star-forming galaxies are indeed the major sources of the reionizing photons
(Silk 2002, private communication). 

\section{Summary}

   The ACS/WFC provides an unprecedented opportunity to probe the LF
of $z\simeq 6$ galaxies at its faint end. The two red filters that it
has, the F775W and the F850LP, are ideal for selecting such objects by
identifying the Lyman-break signature at $6.0\leq z\leq 6.5$. We present our
preliminary analysis on a deep ACS/WFC parallel field.
We discovered a total of 30 plausible candidates to a limit of 28.3
mag in the F850LP-band, 27 of which are to a limit of 28.0 mag. All these
candidates were detected at $S/N\geq 7.2$ in the F850LP-band.

   The contamination to our sample due to either emission-line galaxies or
elliptical galaxies is likely negligible. The major contaminator
would be cool Galactic dwarfs, whose effect is hard to quantify as their spatial
distribution is not well known. To the best of our
knowledge, we estimate that they could contribute at most 4 objects to the
candidate list.

   We derive a galaxy cumulative number density at $6.0\leq z\leq 6.5$ as 
2.3 arcmin$^{-2}$ to a limit of 28.0 mag, which is slightly higher than our
earlier prediction (Yan \etal 2002). While this higher observed value could
be an illusion due to an underestimated contamination of cool Galactic dwarfs,
it could also be real. If the later is the case, one possible explanation is
that the faint-end slope of the actual LF at $z\simeq 6$ may be as steep as
$\alpha=-2$. Given the faintness of these candidates, spectroscopic 
identification is extremely difficult with existing instruments. However, deep
IR imaging data now feasible with the revived NICMOS will enable us to
distinguish these two possibilities. At the very least, the number of
plausible candidates strongly suggests that galaxies with $L<L^*$ do exist in
large number at $z\simeq 6$, and that they are important contributors to the
reionizing background. Future IR surveys with the JWST will allow us to map the
LF beyond $z\simeq 6$ and to even fainter fluxes.

\acknowledgments
The authors would like to thank the anonymous referee for the helpful comments.
We thank Dr. H. J. A. R\"{o}ttgering for carefully reading the manuscript. We
acknowledge funding from NASA/JWST Grant NAG5-12460.


\clearpage

\begin{figure}
\plotone{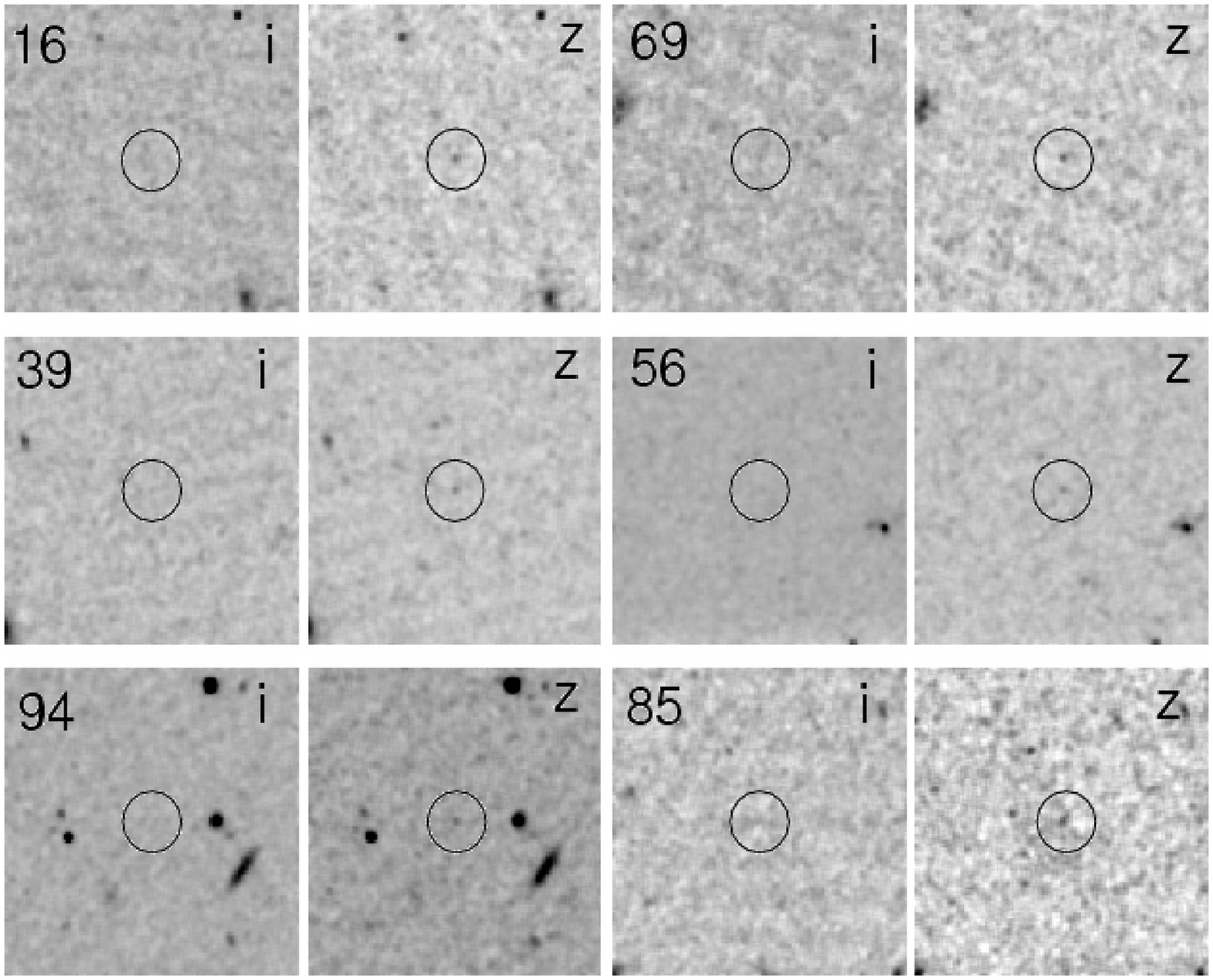}
\caption{The image stamps in the F775W (labeled as $i$) and F850LP (labeled as
$z$) bands for six objects that are randomly chosen from our candidate list.
The images are $7^{'}$ on a side, and have been smoothed by a $5\times 5$
boxcar. The ID's of these candidates are label to the left of the F775W-band
images, and the small circles indicate the locations of the candidates. All the
30 candidates have $S/N\geq 7.2$ in the F850LP band, and their median magnitude
in this band is 27.4 mag.
}
\end{figure}

\begin{figure}
\epsscale{0.8}
\plotone{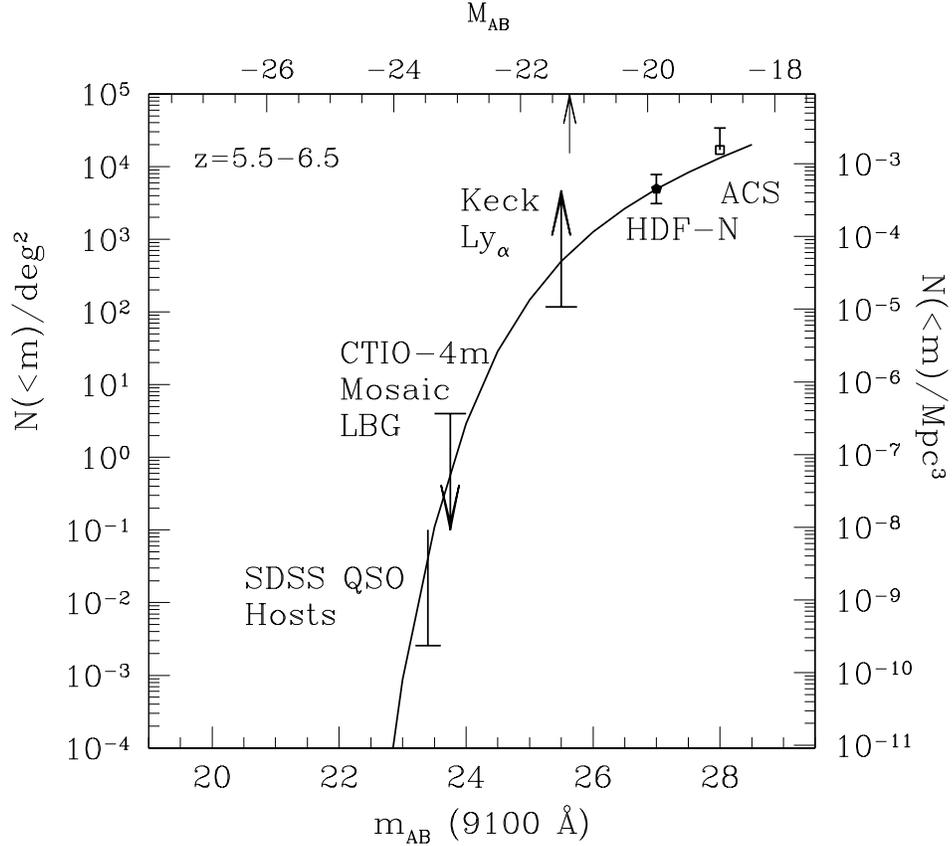}
\caption{The cumulative number density of galaxies at $5.5\leq z\leq 6.5$ 
inferred from the number of $6.0\leq z\leq 6.5$ candidates found in this deep
ACS parallel field is slightly higher than our prediction in Yan \etal
(2002, Figure 1), which was obtained by extrapolating the LF at $z\simeq 3$ to
6 and using the few available data in the HDF-N as the normalization.
For clarity, only the prediction of the high-normalization case for a 
$(\Omega_M,\Omega_\Lambda)=(0.3,0.7)$ universe is reproduced here (the solid
line). The $M^*$ value is marked by an arrow on the absolute magnitude scale
on the top.
This prediction is consistent with nearly all the known constraints at the flux
limit of brighter than 27.0 mag (see Yan \etal 2002 for details). Our current
result (the open square) puts a new constraint at 28.0 mag. The one-sided error
indicates the upper limit that can be inferred from our result by assuming our
candidate search suffers a 50\% incompleteness at the 27.5--28.0 magnitude bin
(see text). If the high observed value is real, it could suggest that the 
faint-end slope of the LF is steeper than $\alpha=-1.6$.
}
\end{figure}

\end{document}